\documentclass[amsmath,aps,prb,twocolumn, preprintnumbers, letterpaper]{revtex4}

\usepackage{graphicx}

\begin{document}

\title{Bayesian estimates of free energies from nonequilibrium work data in the presence of instrument noise.}

\author{Paul Maragakis}
\affiliation{Department of Chemistry and Chemical Biology, 
Harvard University, Cambridge, Massachusetts 02138, USA\footnote{present address: D.~E.~Shaw Research, New York, New York 10036, USA; electronic address: paul.maragakis@deshaw.com}}

\author{Felix Ritort}
\affiliation{Departament de F\'isica Fonamental, Facultat de F\'isica, 
Universitat de Barcelona, 
08028 Barcelona, Spain }
\affiliation{CIBER-BBN, Networking centre on Bioengineering,
Biomaterials and Nanomedicine}

\author{Carlos Bustamante}
\affiliation{Howard Hughes Medical Institute}
\affiliation{Departments of Physics and Molecular \& Cell Biology,
University of California, Berkeley, California, 94720, 
USA}

\author{Martin Karplus}
\affiliation{Department of Chemistry and Chemical Biology, 
Harvard University, Cambridge, Massachusetts 02138, USA}
\affiliation{Laboratoire de Chimie Biophysique, 
Institut de Science et
    d'Ing\'enierie Supramol\'eculaires,
Universit\'e Louis Pasteur, 
F-67083 Strasbourg Cedex, 
France
}

\author{Gavin E. Crooks }
\email{gecrooks@lbl.gov}
\affiliation{Physical Biosciences Division, Lawrence Berkeley National Laboratory, Berkeley, California 94720, USA}

\date{\today}
\preprint{LBNL-62739}

\begin{abstract}
The Jarzynski equality and the fluctuation theorem relate equilibrium free energy differences to non-equilibrium measurements of the work.  These relations extend to single-molecule experiments that have probed the finite-time thermodynamics of proteins and nucleic acids.  The effects of experimental error and  instrument noise have not previously been considered.  Here, we present a Bayesian formalism for estimating free-energy changes from non-equilibrium work measurements that compensates for instrument noise and combines data from multiple driving protocols. We reanalyze a recent set of experiments in which a single RNA hairpin is unfolded and refolded using optical tweezers at three different rates.  Interestingly, the fastest and farthest-from-equilibrium measurements contain the least instrumental noise, and therefore provide a more accurate estimate of the free energies than a few slow,  more noisy,  near-equilibrium measurements.   The methods we propose here will extend the scope of single-molecule experiments; they can be used in the analysis of  data from measurements with AFM, optical, and magnetic tweezers.
\end{abstract}
\maketitle

\renewcommand{\P}{P} 
\newcommand{\F}{\Lambda}
\newcommand{\R}{\tilde{\Lambda}}
\newcommand{\subF}{_{\F}}
\newcommand{\subR}{_{\R}}
\newcommand{\D}{\Delta F}
\newcommand{\DF}{\Delta F\subF}
\newcommand{\DR}{\Delta F\subR}
\newcommand{\W}{W}
\newcommand{\kB}{k_{\mathrm B}}
\newcommand{\kT}{k_{\mathrm B}T}

\nocite{Clausius1865}

\section{Introduction}

A central endeavor of thermodynamics is the measurement of entropy and free energy changes, for which  the principal  experimental methods are  based on the Clausius inequality\cite{Clausius1865}.
One starts with a system equilibrated in one thermodynamic state,~$A$, and then perturbs the system, following some explicit protocol, until the control parameter corresponds to a new thermodynamic state,~$B$.  If the temperature~$T$ of the surroundings is fixed, the change in entropy, $\Delta S=S_B -S_A$, is related to the flow of heat $Q$ into the system:
\begin{equation}
	\Delta S \geq \beta \langle Q \rangle,
\label{DS}
\end{equation}
where $\beta=1/\kB T$, and $\kB$ is the Boltzmann constant.
Equivalently, the free energy difference $\Delta F =F_B - F_A= \Delta \langle U \rangle - \Delta S/\beta$ is related to the work $W$ done on the system:  
\begin{equation}
	\Delta F \leq \langle W\rangle .
\label{DF}
\end{equation}
Here we use the sign convention $\Delta U = Q + W$. The angle brackets indicate an average over many repetitions of the same experiment. In macroscopic systems individual observations do not differ significantly from the mean.  But for a microscopic system the fluctuations from the mean can be large and the inequality only holds on average (i.e.,\  not for individual measurements). 

\begin{figure}[t]
\includegraphics{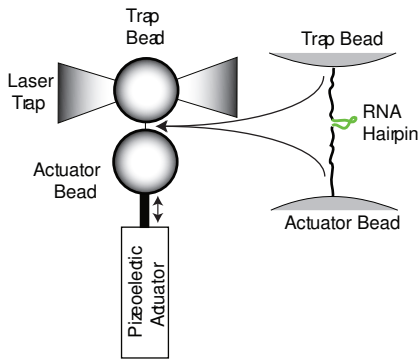} 
\caption{Non-equilibrium work measurements for folding and unfolding an RNA hairpin\cite{Collin2005}.  A single RNA molecule is attached between two beads via hybrid DNA/RNA linkers. One bead is captured in an optical laser trap that can measure the applied force on the bead. The other bead is attached to a piezoelectric actuator, which is used to irreversibly unfold and refold the hairpin\cite{Hummer2001a,Collin2005,Bustamante2005, Hummer2005, Dhar2005,Imparato2005a,Ritort2005,Ritort2006a,Braun2004a, Braun2004b}.
}
\label{fig-exp}
\end{figure}

It was recently discovered that equilibrium free energy differences can also be determined by measuring the work performed during irreversible transformations, using the Jarzynski\cite{Jarzynski1997b,Jarzynski1997a,Jarzynski1998,Crooks1998} and work fluctuation relations\cite{Crooks1999a,Crooks2000}. These theoretical insights have  been used to determine the unfolding free energy of an RNA hairpin\cite{Hummer2001a,Liphardt2002,Bustamante2005,Collin2005,Ritort2006a} from finite-time, non-equilibrium experiments, as described in Fig~\ref{fig-exp}.
We consider a protocol (labeled $\Lambda$) that starts with an equilibrated system, and then transforms an external control parameter  from an initial value $A$, to a final value $B$ in a finite time. (In the RNA hairpin unfolding experiments, the control parameter is the distance between the center of the optical trap and the center of the fixed bead.) This perturbation drives the system out-of-equilibrium. Once the protocol ends, the control parameter is again fixed, and the system can relax back to thermal equilibrium. One can also run the protocol in reverse, starting with a system equilibrated with the control parameter at~$B$, and then transform the system  
through the reverse sequence of intermediate control parameters,
to~$A$. We label this conjugate protocol~$\R$. Due to the reversibility of the microscopic dynamics, the probability~$\P(\W|\DF, \F)$ of measuring a particular value of the work during protocol $\F$ is related to the work probability density of the conjugate protocol, $\R$, by the following work fluctuation symmetry\cite{Crooks1999a,Crooks2000,Blau2002,Evans2002a,Evans2003, Reid2005, Bustamante2005,Collin2005}: 
\begin{equation}
\frac{ \P(+\W| \DF, \F )}{\P(-\W| \DR, \R)} = e^{+\beta \W- \beta \DF},
\label{WFT}
\end{equation}
with $\DF$ ($=-\DR$) the change in free energy associated with
the change of the external control  parameter in protocol $\F$ ($\R$).
This relation immediately implies the Jarzynski equality\cite{Jarzynski1997a,Jarzynski1997b,Jarzynski1998,Crooks1998,Crooks1999a,Jarzynski2004b}
\begin{eqnarray}
\left \langle e^{-\beta W} \right\rangle &=& \int dW  \P(+\W| \DF, \F ) \,e^{-\beta W} \nonumber \\
&=& \int dW {\P(-\W| \DR, \R)} \,e^{- \beta \Delta F} \nonumber \\
&=&e^{-\beta \Delta F}.
\label{eq:jarzynski}
\end{eqnarray}
In other words, a Boltzmann weighted average of the irreversible work recovers the equilibrium free energy difference from a non-equilibrium transformation.
The Clausius relation follows by an application of Jensen's inequality, $\ln \langle \exp(x)\rangle \geq \langle x \rangle$. 

\begin{figure}[t]
\includegraphics{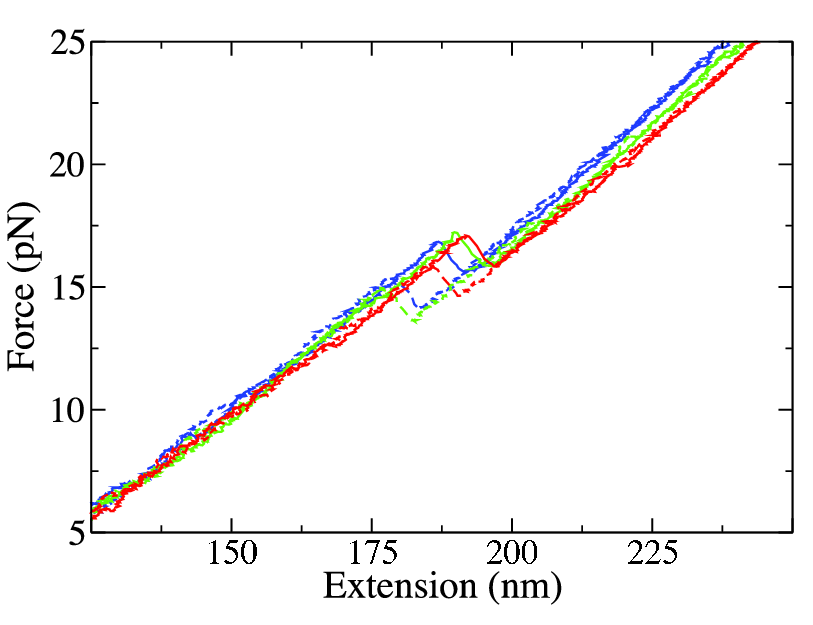} 
\caption{Typical force extension curves in the unfolding (solid
lines) and folding (dashed lines) of a 20 base pairs
RNA hairpin. Different colors correspond to different
unfolding-folding cycles. The rip in force observed
around 15pN corresponds to the cooperative
unfolding/folding transition. The area below the
force-extension curve is equal to the mechanical work
done on the RNA hairpin.  Because the transformations are irreversible, the work performed varies from one unfolding or refolding measurement to the next.  Drift effects observed in force extension curves arise from different causes, including air currents, mechanical vibrations and temperature changes.
   }
 \end{figure}

Given the thermodynamics preamble, we can rephrase the problem of measuring the free energy as follows: How do we calculate the most accurate, least biased, estimate of the free energy, given a finite number of irreversible work measurements?\cite{Crooks2000, Hendrix2001, Hummer2001a,Hummer2001b, Hummer2002, Zuckerman2002, Shirts2003, Park2003, Gore2003, Sun2003, Wu2004a,Ytreberg2004, Wu2005, deKoning2005, Lechner2006,Jarzynski2006a}  We consider both the statistical error due to limited data and, for real experiments, the additional error due to measurement noise. Furthermore, we may wish to simultaneously  combine the data from multiple protocols connecting the same thermodynamic  states~\cite{Maragakis2006}. For example, in the single-molecule experiment described in Fig.~\ref{fig-exp}, the same RNA hairpin was unfolded at three different rates, with each dataset providing a different compromise between statistical and experimental errors.

The Clausius relations are exact equalities only for infinitely slow, thermodynamically reversible transformations, where the irreversible dissipation is zero. A transformation that occurs in a finite time provides only an upper bound to the free energy  and a lower bound to the entropy change. (Since entropy and free energy are state variables, the reverse transformation, from thermodynamic state $B$ back to $A$, provides a lower (upper) bound to the same  free energy (entropy) change.)
One approach to analyzing irreversible transformations is to directly apply the Jarzynski relation~\cite{Jarzynski1997a,Hummer2001a,Liphardt2002,Douarche2005a}. However, this identity strictly holds only in the limit of an infinite number of  repeated experiments.  For  a finite number of measurements, we again obtain an inequality that only holds on average~\cite{Jarzynski1997b}, and the free energy estimates tend to be strongly biased~\cite{Hummer2001b, Gore2003,Ytreberg2004, Wu2004a, Imparato2005c, Kofke2006, Jarzynski2006a,Lua2005a,Crooks2007a}.  Because the magnitude of the bias depends on the protocol, one cannot  reliably combine data from different protocols\cite{Minh2006a}. Moreover, the Jarzynski relation is sensitive to measurement noise 
and variations in the experimental setup (e.g., heterogeneity in the attachments and variable length  of tethers).  Broadening of the work distribution leads to a bias in the estimated free energy,
since smaller work values contribute more than larger work values in the exponential average of Eq.~\ref{eq:jarzynski}.

Bennett laid the foundations for the solution to this problem in his development of the acceptance ratio method for free energy perturbation calculations~\cite{Bennett1976} (a technique for computing free energy changes by simulating infinitely fast transformations).  He realized that an optimal solution requires combination of work measurements from both forward and reverse switches.  The acceptance ratio method was 
later
extended to finite-time switches~\cite{Crooks2000}, shown to a maximum-likelihood free energy  method~\cite{Shirts2003,Shirts2005}, related to the problem of logistic regression~\cite{Shirts2003,Maragakis2006,Anderson1972,Gelman2004}, and  
extended to a network of thermodynamic states connected with many protocols~\cite{Maragakis2006}.
In this paper, we develop a Bayesian formalism that extends  these results to provide not only a reliable estimate of the free energy, but also reliable estimates of the statistical uncertainty. In this formalism, it is straightforward to incorporate additional prior information about the experiment into the analysis.  In particular, we show how to allow for experimental measurement noise. The magnitude of the noise can be determined from the data and an error-corrected free energy estimate recovered.  We  use this approach to reanalyze a recent experiment in which a single RNA hairpin was unfolded and refolded  at three different rates using optical tweezers~\cite{Maragakis2006}.

\section{Posterior Free Energy Estimate}

Formally, we require the probability that the free energy change $\D$ has a particular value, given a collection of work measurements $\W$, the protocol used for each measurement (either ${ \F}$ or  $\R$), and the (fixed)  temperature of the environment $T$.  Initially, we consider the simplest case, in which there are two protocols that are conjugate to each other, so that  the work distributions are related by the fluctuation relation~Eq.~(\ref{WFT}).  We also  assume, for now, that the measurements are error free.

The essential element in solving this problem is to treat both the work and the protocol as random variables that are uncorrelated from one observation to the next\cite{Shirts2003}. We rewrite the free energy probability density given a single measurement in terms of these variables using  Bayes'  rules, $\P(A|B) = \P(B|A)\P(A)/\P(B)$:
\begin{equation}
\P(\DF | W, \F) = \frac{\P(W, \F| \DF) \P(\DF) }{\P(W, \F)}.
\end{equation}
Since {\it a priori} the free energy could be positive or negative and of any magnitude, the prior distribution of free energy $\P(\DF)$ can be reasonably taken as uniform 
(see Kass and Wasserman\cite{Kass1996} for an in-depth discussion of priors).
The denominator, which does not depend on $\DF$, can be absorbed into a normalization constant. 

The distribution $\P(W, \F| \DF)$ is the final undetermined factor on the right-hand side of Eq.~(5).
In the absence of detailed knowledge about the work  likelihood for the system under investigation, we should  choose a maximally uninformative, system independent distribution.  If the work were not conditional on the free energy we could again assign a uniform distribution, since a single work measurement could be positive or negative and of any magnitude. But,  
we expect
 that the work will probably  (but not certainly) be larger than that value of the free energy. Concretely, any work probability distribution must satisfy the work fluctuation symmetry, Eq.~(\ref{WFT}). We can satisfy this constraint by first considering the symmetrized distribution $\P(W, \F| \DF) + \P(-W, \R| \DR)$. This averaged distribution does not need to satisfy any symmetry and therefore we can again assign a maximally uninformative improper prior:
\begin{equation}
\P(W, \F| \DF) + \P(-W, \R| \DR) = \mbox{constant}.
\end{equation}
However, the work fluctuation relation implies that
\begin{equation}
\frac{\P(+W, \F| \DF)}{\P(-W, \R| \DR)} = e^{\beta \W- \beta \DF + M_{\F}}
\end{equation}
where $M_{\F}=\ln \P(\F|\DF)/\P(\R|\DR)$. It follows that 
\begin{equation}
\P(W, \F| \DF) \propto \frac{1}{1 + e^{\beta \W- \beta \DF + M_{\F}} } 
\end{equation}
Together with an uninformative free energy prior, we finally obtain
\begin{eqnarray}
\P(\DF| W, \F) &\propto& \P(W, \F| \DF) \nonumber \\
&\propto&  f\big( \beta \W - \beta \DF+ M_{\F}\big),
\label{single} 
\end{eqnarray}
where  $f(x)$ is the logistic function (Fig.~\ref{fig-logistic}), the cumulative distribution function of the standard logistic distribution (see appendix, Fig.~\ref{fig-approx-gaussian}):
\begin{equation}
f(x) =\frac{1}{1 + e^{-x}}.
\end{equation}

\begin{figure}[t]
\includegraphics{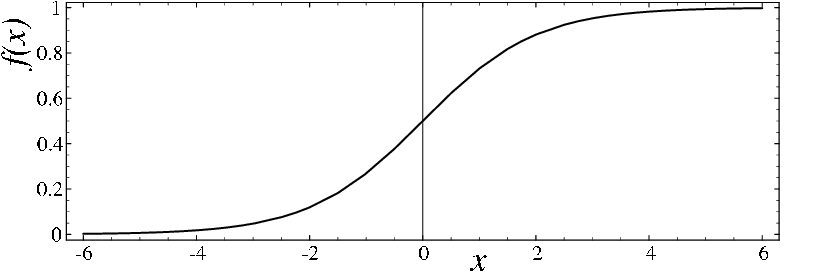} 
\caption{The standard logistic function, $f(x)= 1/ (1+e^{-x})$.}
\label{fig-logistic}
\end{figure}

\begin{figure}[t]
\includegraphics{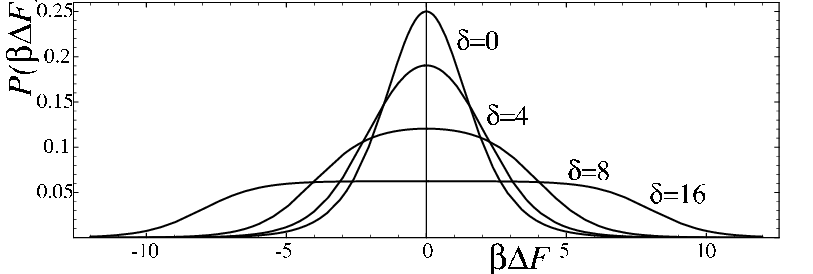} 
\caption{Posterior free energy given  two work  measurements, one from each of two conjugate protocols with values $\beta W=\pm\frac{1}{2}\delta$. The posterior variance, $\pi^2/3 + \delta^2/12$, is minimized when the rectified work variables coincide, and increases quadratically with separation. 
}
\label{fig-post}
\end{figure}

Essentially, each measurement of the work provides a soft upper bound to the free energy change. Measurements made on the conjugate protocol provide soft lower bounds to the same free energy.  Therefore, combining measurements from conjugate protocol pairs provides reliable,  but fuzzy, free energy bounds. This is in contrast to the Clausius inequality [Eq.~(\ref{DF})] where the {\it average} work provides a hard bound to the free energy change.

Figure~\ref{fig-post} illustrates the posterior distribution resulting from combining  two work measurements,  one from each of a conjugate protocol pair, where the measured values are $\beta W=\pm\frac{1}{2}\delta$. If the work values are widely separated, then the posterior free energy distribution is broad and flat. We only obtain a tight constraint on the free energy if the separation is less than about $4\kT$. The minimum uncertainty for a single pair of measurements  is $\sigma\approx 1.8 \kT$, which  occurs when $\delta=0$.

Assuming that each measurement of the work is independent, we can combine measurements by multiplying the separate posterior distributions together. 
So far, we have been considering a single pair of conjugate protocols switching between two thermodynamics states. However, it was recently demonstrated that we can combine measurements from many different protocols connecting many different thermodynamic states in a network of transformations\cite{Maragakis2006}. Each measurement provides a single soft constraint [Eq.~(\ref{single})], which we can combine by multiplying the different posterior distributions:
\begin{equation}
\P( {\mathbf F}  | {\mathbf \W},{\mathbf \F}) 
= \frac{1}{\mathcal C}  \prod_{k=1}^{N} f\big( \beta \W_k- \beta \D_{\F_k} + M_{\F_k}\big) ,
\label{estimate}
\end{equation}
where $W_k$ is the work measured in the $k$th experiment, performed with  protocol $\Lambda_k$,  $\Delta F_{\F_k}$ is the free energy  change associated with that protocol, $\mathcal C$ is a normalization constant and $N$ is the total number of measurements.  In the simplest case we have only a single conjugate protocol pair, forward and reverse.  In general, we can have many different protocols (for example, pulling a molecule apart at different loading rates.), and different protocols could connect different thermodynamic states~\cite{Maragakis2006}. In the equation above, ${\mathbf F} = \{F_1, F_2, F_3, \ldots \}$ are the free energies of the initial and final states of these transformations. At least one free energy $F_i$ is fixed at zero, or some other convenient reference point, since only differences in free energy are significant.

The $M_{\Lambda_k}$ terms compensate for a difference in the probability of observing a forward or reverse protocol from a conjugate protocol pair. In the absence of detailed prior information about the work distributions, it is best to pick each member of a conjugate pair equally often\cite{Bennett1976}. However, the difficulties of  real world experiments may result in unequal numbers of forward and reverse measurements. In such  cases,  we can estimate a reasonable value for $M_{\Lambda_k}$ from the number of observations, $N_\F$, obtained from each protocol:
\begin{equation}
 M_{\Lambda}=\ln { \frac{\P(\F|\DF)}{\P(\R|\DR)} } \approx  \ln { \frac{N\subF+1}{N\subR+1} }.
\label{M}
\end{equation} 
The additional `$+1$' is a pseudocount which regularizes the frequency estimate. It can be justified as a Laplace prior on the probabilities\cite{Jaynes2003,Durbin1998}. Note that without this regularization, Eq.~(\ref{M}), and thus also Eq.~(\ref{estimate}), would become invalid in the single sample limit.  With the addition of the pseudocount, the probability distribution  in Eq.~(\ref{estimate}) may still only produce one-sided bounds (for example, when there is no protocol that ends in a certain state, one has at best an upper bound for the free energy of that state.) 
 However, we could recover a finite  free energy posterior distribution
if we were to use a more informative free energy prior in Eq.~(\ref{estimate}).

The experimental measurements of the work values can typically be considered to be uncorrelated.  However, when the measurements, or simulation results, are correlated, the maximum likelihood, or Bayesian estimates, may need to be modified to result in an optimal estimate of the free energy\cite{Gelman2004}.  
In the absence of a general-purpose formulation for correlated work measurements, the estimators discussed in this paper are likely to underestimate the errors. 

The Bayesian free energy posterior is an optimal estimate in the sense that it uses all of the available data and makes the fewest possible assumptions. 
 We can, in principle,
improve the estimate by incorporating additional information, either by using more informative priors, or by adding additional assumptions, for example, by assuming that the work distribution is smoothly varying\cite{Bennett1976}, or that it can be parameterized in terms of a particular functional form\cite{Nanda2005}.

In many practical cases, the posterior distribution of $\D$ quickly converges to a normal one as a consequence of the central limit theorem.  We can summarize this posterior distribution with a point estimate and reasonable error bounds, for example the posterior mean free energy and 95\%  confidence intervals.  The posterior mean will coincide with the maximum likelihood, and the confidence interval will be $\pm2$ standard deviations.

\section{Experimental Errors}

\begin{figure}[t] 
(a)$\qquad\qquad\qquad\qquad\qquad\qquad\qquad\qquad\qquad\qquad\qquad\qquad$
(b)\includegraphics{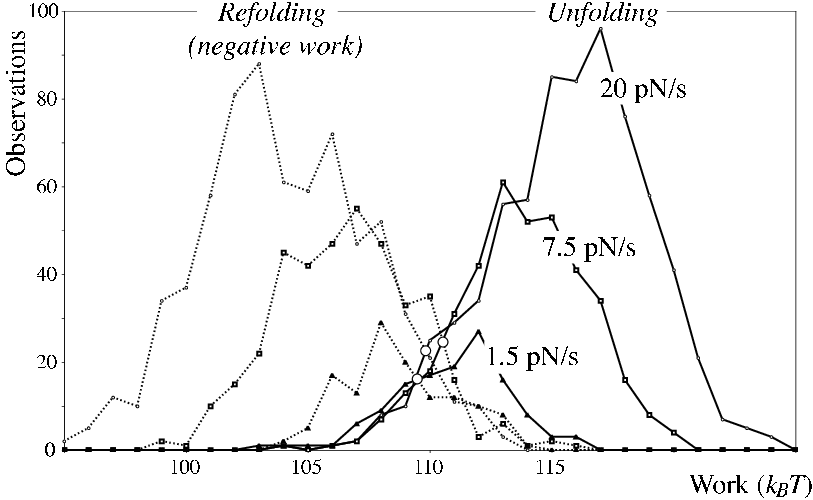} 
(c)\includegraphics{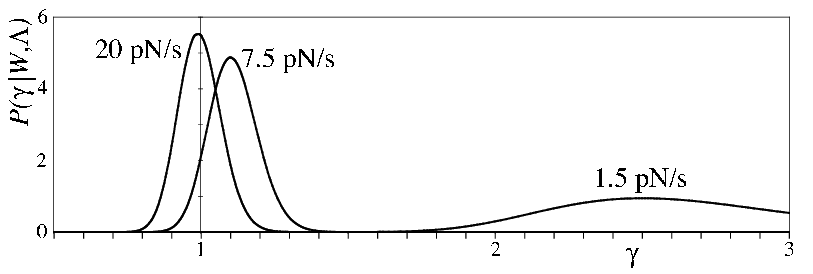} 
\includegraphics{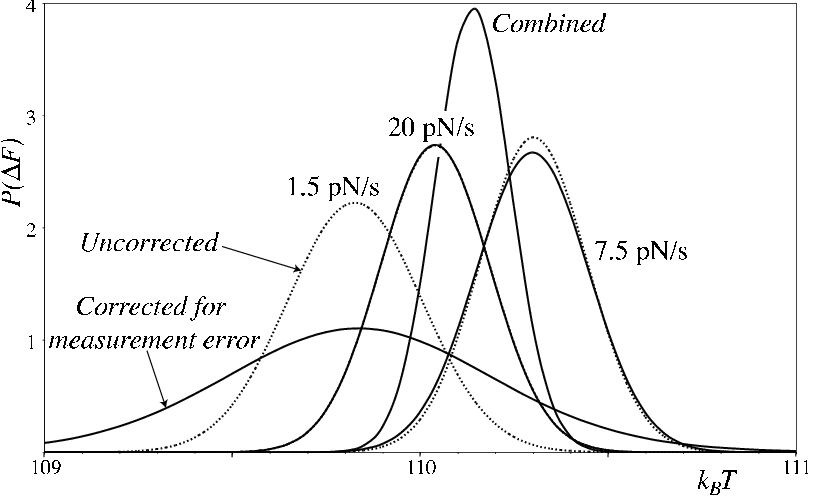} 
\caption{ 
(a) Histograms of work measurements for folding and unfolding an RNA hairpin at three different rates. Observations are binned into integers centered at 1 $\kB T$ intervals. This data corresponds to Fig.~2 of Collin {\it et al.}\cite{Collin2005}. Note that Eq.~(\ref{WFT}) predicts that the folding and unfolding work distributions cross at the free energy change. 
(b) The posterior distribution of the error correction factor $\gamma$ [Eq.~(\ref{eq-final})]. 
(c) Posterior free energy derived from  the data in (a), both with [Solid line, Eq.~(\ref{eq-final})]  and without [Dashed line, Eq.~(\ref{estimate})] correction for measurement noise. Notice that the correction is substantial for the slowest experiment (1.5 pN/s), minor for the intermediate rate, and the corrected and uncorrected posteriors are indistinguishable (at this scale) for the fastest rate. The most reliable free energy estimate is obtained by combining the three separate noise corrected free energy posterior distributions.
}
\label{fig-results}
\end{figure}

 The preceding analysis 
 does not include 
 the possibility of experimental 
 errors, an omission that we now address, since real experiments are not ideal and real measurements  can be inaccurate.

We initially assume that the instrument error can be adequately described as additive white noise 
with zero mean and standard deviation~$\sigma$. Since we do  not know the magnitude of the noise,  we estimate the joint distribution of the free energy and the noise, then integrate out the noise to obtain a final free energy estimate:
\begin{equation}
\P(\Delta F_{\Lambda} |  \W, \F) = \int \P(\Delta F_{\Lambda} , \sigma| \W, \F)   d\sigma.
\end{equation}
Let us write $W = w+\epsilon$ where $W$ is the observed work value, $w$ is the true work and 
$\epsilon$ is the measurement error.  Using Eq.~(\ref{single}) we get,
\begin{eqnarray}
\lefteqn{\P(\DF, \sigma | W, \F) \propto} \qquad\\ &&  
\int\limits_{-\infty}^{+\infty}  
f\big(  \beta W-\beta\epsilon  - \beta\DF  + M_\F) \,{\mathcal N}(\epsilon; 0,\sigma) \;d\epsilon.
 \nonumber
\end{eqnarray}
Here, ${\mathcal N}(x; \mu,\sigma)$ is a Gaussian distribution with mean $\mu$ and standard deviation $\sigma$. [See Eq.~(\ref{gaussian})].

\begin{figure}[t]
\includegraphics{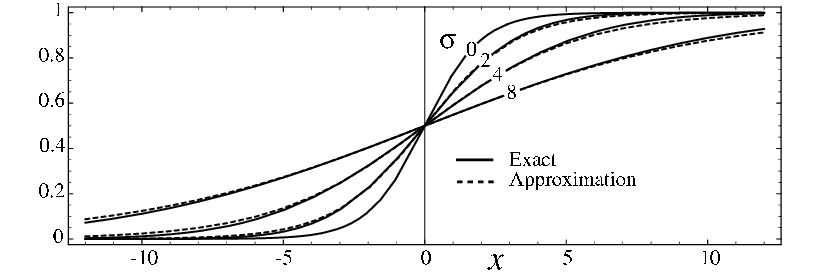} 
\caption{ The approximation of the sigmoidal  function $g(x; \alpha, \sigma)$ [Eq.~(\ref{eq_g})] by the logistic  function $f(x; \gamma) = 1/(1+\exp(-x/\gamma))$, where $\gamma=\sqrt{ 1+ \pi\sigma^2/8}$ [Eq.~(\ref{eq_gamma})]. The absolute difference between the functions is always less than 0.02. }
\label{fig-approx}
\end{figure}

This convolution of a logistic function and a Gaussian distribution generates a new sigmoidal function, illustrated in Fig.~\ref{fig-approx}. This function does not have a simple closed form, but fortunately it can be closely  approximated by a reparametrized logistic distribution 
\begin{equation}
\P(\DF, \sigma | W, \F) \propto 
f\Big( \frac{1}{\gamma}( \beta W - \beta\DF  + M_\F) \Big) ,
\end{equation}
where the parameter $\gamma = \sqrt{1+ \pi \beta^2 \sigma^2/8 }$  essentially  acts as a correction factor to the work fluctuation symmetry. (The mathematical details are given in the appendix.)

Having proceeded this far, we no longer need to assume that the errors are a result of white noise. Instead, we will treat  $\gamma$ as the principle experimental error factor directly, without reference to an explicit error model or to the standard deviation of the noise, $\sigma$. For example, a systematic miscalibration of the work measurement or an incorrect thermostat would also result in a non-unit $\gamma$. In such cases $\gamma$ could be less than 1.  Therefore, we allow~$\gamma$ to be any positive number. We introduce an uninformative prior for $\gamma$, $\P(\gamma) = 1/\gamma$.  This distribution is scale invariant and follows given only that $\gamma$ is positive and {\it a priori} of unknown magnitude\cite{Jaynes2003}. 
We can now average over the free energy to obtain the posterior distribution of the error correction factor $\gamma$, or average over the error correction factor to obtain the posterior free energy estimate corrected for instrument error 
\begin{eqnarray}
\lefteqn{\P({\DF} | {\W},{\F}) =}\qquad 
\label{eq-final} \\ &&
 \frac{1}{\mathcal C'} \int\limits_0^{+\infty} \frac{1}{\gamma} \prod_k 
f\Big( \frac{1}{\gamma}( \beta \W_k - \beta\D_{\F}   + M_{\F}) \Big) 
d\gamma,
\nonumber 
\end{eqnarray}
where ${\mathcal C'}$ a normalization constant.
Note that  instrument error, and thus the distribution of~$\gamma$, 
will vary with the protocol. 
One could construct a complex hierarchical prior\cite{Gelman2004} for
the experimental error factors, that would feed information about the 
typical scale of the errors from one protocol to the next.  In this work, we find it sufficient to  
estimate $\gamma$ independently for each protocol, and obtain a final 
posterior:
\begin{eqnarray}
\P({\mathbf F} | {\mathbf \W},{\mathbf \F}) =
\prod_\F 
\P({\DF} | {\W},{\F}).
\end{eqnarray}
Here, as in Eq.~(\ref{estimate}), ${\mathbf F} = \{F_1, F_2, F_3, \ldots \}$ are the free energies of the initial and final thermodynamic states.

Another potential source of errors
arises from
 unintended variations of the experimental procedure from one measurement to the next. 
For example, we may intend to  forcibly unfold  an RNA hairpin in a particular time, but  each experimental run may be slightly faster or slower than another. Instead of an experiment being described by a single protocol, each measurement is made with a similar,  but slightly different procedure (e.g. due to hysteresis effects in the mechanical response of the actuators).  However, if a protocol variation has the same probability both forward and reverse, then the factor $M_{\Lambda}$ [Eq.~(\ref{M})] does not change. Consequently, if the variations in protocol are statistically  the same for the conjugate forward and reverse protocol pairs then that variation has no effect on the free energy estimate.

\begin{table}[t]
\begin{tabular}{rccccc}
& $N_U$ & $N_R$ & $\Delta F$  &  $\Delta F$    & $\gamma$ \\ 
& & & {\scriptsize (Uncorrected)}&  {\scriptsize (Corrected)}&\\\hline
1.5 pN/s &127& 129 & $109.8\pm0.4$ &  $109.8 \pm0.8$ & $2.70 \pm 1.00$\\
7.5 pN/s  & 384 & 383 & $110.3\pm 0.3$ & $110.3\pm 0.3$  &$1.11\pm0.17$  \\
20 pN/s  & 699 & 696 & $110.0 \pm 0.3$ & $110.0\pm 0.3$  & $1.00 \pm 0.14$ \\
Combined  & &         &                                   & $110.1 \pm 0.2$& 
\end{tabular}
\caption{ Summary of results graphed in Fig.~(\ref{fig-results}). $N_U$ and $N_R$: Number of
unfolding and refolding work measurements at each pulling rate, respectively. $\Delta F$: Posterior mean free energy estimate with 95\% confidence intervals, both corrected and uncorrected for measurement error. $\gamma$: Posterior mean estimate of the noise correction factor, with 95\% confidence intervals 
}
\label{table-results}
\end{table}

\section{Application and Discussion}

Figure~\ref{fig-results} shows the result of applying the Bayesian free energy estimate to data from the single-molecule RNA pulling experiments  reported in \cite{Collin2005}, both with and without noise correction. This data set is particularly
useful to illustrate the previous analysis, 
since it represents three distinct protocols; 
i.e.\
the same RNA hairpin is unfolded at three different rates: slow, medium, and fast. The free energy change is the same in each case; we can see that this is qualitatively true by noting that the forward-reverse work histograms all cross at roughly the same  value of the work. The experimental noise is expected to accumulate during a single experiment, and  so we expect the data from the fastest pulling rate to be contaminated with the least measurement error. This is indeed what the Bayesian error analysis finds: $\gamma$ approaches $1$ as the pulling rate increases.

Qualitatively, the effect of instrument noise is to broaden both the forward and reverse work distributions. 
This broadening tends not to significantly change the crossing point, but it does increase the overlap between the conjugate distributions.  Therefore, 
ironically, 
the instrument error does not greatly change the free energy estimate, but it does significantly (and erroneously) reduce the calculated error bars.  Fortunately, the noise invalidates the fluctuation theorem, and the magnitude of that violation allows us to estimate the magnitude of the instrument errors and to extract noise-corrected free energy estimates with meaningful error bounds. 

A useful feature of this error analysis is that we can use the parameter $\gamma$ as a measure of how well the experiments have confirmed the work fluctuation relation [Eq.~(\ref{WFT})]. For the fastest pulling, highest quality data, we find that $\gamma=1\pm0.14$;  in other words, the fluctuation relation is confirmed to within 14\% at the 95\% confidence limit. Although more accurate constraints can be obtained by performing experiments on systems with simple potentials\cite{Carberry2004a, Douarche2005a,Schuler2005,Wang2002, Wang2005a,Wang2005b}, this is the best available experimental data for irreversibly switching a complex system\cite{Trepagnier2004,Collin2005}.
 We can also use the interrelation between the noise and the correction factor ($\gamma = \sqrt{1+ \pi \beta^2 \sigma^2/8 }$)  to estimate the measurement accuracy needed to improve this result. For example,  if we wish to confirm the fluctuation relation to better than 1\%,  then the work must be measured to better than $\approx  \frac{1}{4} \kT$ accuracy, which is well within the limits of modern optical tweezer instruments.

The quantitative effect of the noise  corrections to $\Delta F$ can be seen in Fig.~\ref{fig-results}c and  Table~\ref{table-results}. The noise correction makes a substantial difference to the free energy confidence interval for the slowest data, but very little difference to the posterior mean free energy or the error bounds for the faster data. Note that the free energy considered in this analysis includes unfolding the RNA hairpin and stretching the DNA/RNA handles; deconvoluting the contributions of the handles introduces additional uncertainty not considered here\cite{Hummer2001a,Liphardt2002,Collin2005}. 
Having applied the  instrument noise correction, we can safely combine the posterior free energy estimates from the three different protocols to obtain a combined estimate of $\Delta F = 110.1\pm0.2\kT$. This result is a substantial improvement over the best, single protocol, maximum likelihood estimate,  $\Delta F = 110.2\pm0.6\kT$, extracted from the same data\cite{Collin2005}.

In summary, we have presented a Bayesian formalism for estimating free-energy changes from non-equilibrium work measurements.  The formalism compensates for instrument noise and combines results from multiple experimental protocols.  The method is widely applicable and could be used in the analysis of single-molecule experimental data from optical tweezers, AFM, or magnetic tweezers.
Together with advances in single-molecule traps and use of multiple experimental setups (e.g., changing bead sizes, trap power, or the length of the handles), it will aid in extending the scope of single-molecule experiments.

\begin{acknowledgments}
This research was supported by the U.S. Dept of Energy, under contracts
DE-AC02-05CH11231.   The research of F.R.\  was supported by the Spanish and Catalan
research councils
FIS2004-3454, NAN2004-09348, and SGR05-00688. The research of C.B.\ was supported
by NIH Grant GM 32543 and U.S.\ Dept.\ of Energy grant AC0376Sf00098.
The research of M.K. at Harvard was supported in part by a grant from the NIH. 
\end{acknowledgments}


\section*{Appendix : Approximate convolution of a logistic function with a Gaussian distribution}

\begin{figure}[t]
\includegraphics{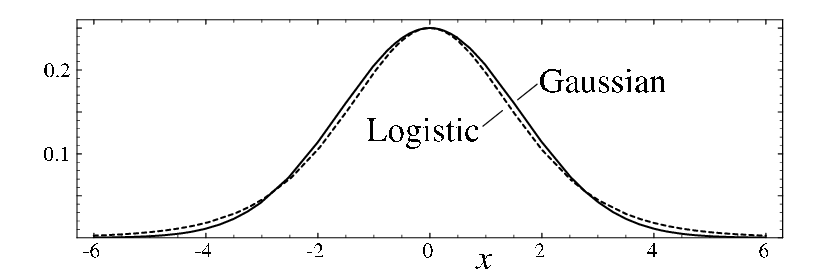} 
\caption{ The approximation of the standard logistic distribution by the Gaussian distribution with zero mean and standard deviation $\sqrt{8/\pi}$. }
\label{fig-approx-gaussian}
\end{figure}

We are interested in the function
\begin{equation}
g(x; \alpha, \sigma) = \int_{-\infty}^{+\infty} f(x+\epsilon; \alpha)\, {\mathcal N}(\epsilon; 0, \sigma) d\epsilon 
\,,
\label{eq_g}
\end{equation}
the convolution of  a logistic (or Fermi) function
\begin{equation}
f(x; \alpha) = \frac{1}{1+\mathrm{e}^{- x/\alpha}}
= \frac{1}{2} + \frac{1}{2} \tanh{ \frac{x/\alpha}{2} },
\label{eq_f}
\end{equation}
 with a Gaussian (or normal) distribution with zero mean and standard 
deviation $\sigma$:
\begin{equation}
{\mathcal N}(x; \mu, \sigma) = \frac{1} {\sqrt{2 \pi \sigma^2}  }\exp\left(-\frac{(x-\mu)^2}{2\sigma^2}\right).
\label{gaussian}
\end{equation}
The function $g(x; \alpha, \sigma)$ does not have a simple, closed form. However, as is illustrated in the figure, it can be  reasonably approximated by a reparameterized logistic function:
\begin{equation}
 g(x; \alpha, \sigma) \approx f(x; \gamma),
\end{equation}
 where $\gamma$ is a function of $\alpha$ and $\sigma$.  
 We fix $\gamma$ by  requiring equality of the derivative at the origin, since,
for our purposes, it is more important to minimize the errors around the origin than elsewhere. The value of 
 $g(x; \alpha, \sigma)$ at the origin is $1/2$, the same as $f(0;\gamma)$. 
Note that
\begin{equation}
 \left.\frac{d}{dx} f(x;\gamma) \right|_{x=0} 
 = \left. \frac{1}{2\gamma+2\gamma \cosh{\left(x/\gamma \right)} } \right|_{x=0}
 = \frac{1}{4\gamma},
\end{equation}
and therefore
\begin{eqnarray}
\gamma^{-1}
&=& 
4  \left.\frac{d}{dx} g(x;\alpha,\sigma) \right|_{x=0}  
 \nonumber\\ &=&
4 \int_{-\infty}^{+\infty} \left(\left. \frac{d}{dx}  f(x+\epsilon; \alpha) \right|_{x=0}\right) {\mathcal N}(\epsilon; \sigma) d\epsilon 
 \nonumber \\ &=&
  4 \int_{-\infty}^{+\infty}  \left( \frac{1}{2\alpha+2\alpha\cosh{\epsilon/\alpha}} \right) {\mathcal N}(\epsilon; \sigma) d\epsilon .
\end{eqnarray}
The expression inside the bracket is a logistic distribution, which is closely approximated by the Gaussian 
distribution  ${\mathcal N}(\epsilon; 0, \alpha \sqrt{8/\pi})$ (See Fig.~\ref{fig-approx-gaussian}).  
These parameters ensure that the two distributions agree exactly at the origin.
%
%
Therefore, our problem  reduces to a straightforward Gaussian integral:
\begin{eqnarray}
\gamma^{-1}
&\approx&
4 \int_{-\infty}^{+\infty}  {\mathcal N}(\epsilon; 0, \alpha\sqrt{\frac{8}{\pi} } ) \,{\mathcal N}(\epsilon; 0, \sigma) d\epsilon  
\nonumber \\ 
 \gamma
 &=&  \sqrt{ 1+ \frac{\pi}{8\alpha^2} \sigma^2}.
 \label{eq_gamma}
\end{eqnarray}
 For $\alpha=-1/\beta$  we recover the case of white noise discussed in the main text.

\bibliography{GECLibrary}

\end{document}